\documentstyle[aps,twocolumn,epsf]{revtex}
\begin{document}

%

\title{Time Correlation Functions of Three Classical
 Heisenberg Spins on an Isosceles Triangle and on a Chain}
\author{Marco Ameduri and Richard A.\ Klemm}
\address{Max-Planck-Institut f\"{u}r Physik komplexer Systeme\\
 N\"{o}thnitzer Str.\ 38, D--01187 Dresden, Germany}

\maketitle

\begin{abstract}
  At arbitrary temperature $T$, we solve for the  
dynamics of single molecule
magnets composed of three 
classical Heisenberg spins either on a chain with two equal exchange
 constants $J_1$,    or on an isosceles triangle
with a third, different exchange constant $J_2$. 
As $T\rightarrow\infty$,
  the Fourier transforms and long-time asymptotic behaviors 
of the two-spin time correlation functions
 are evaluated exactly.
 The lack of  translational symmetry on a chain or 
an isosceles triangle   yields time 
correlation functions  that differ strikingly 
from those on an equilateral triangle  with $J_1=J_2$.
 At low $T$, the Fourier transforms of the two autocorrelation functions
with $J_1\ne J_2$ show one and four  modes, respectively.
For a semi-infinite $J_2/J_1$ range, one mode is a central peak. At the origin of
this range, this mode has a novel
scaling form.

\end{abstract}


%
%

\section{Introduction}

 Recently, there has been   a substantial  interest in the physics of magnetic
 molecules, or single molecule magnets
 (SMM's). \cite{exp1,exp2,exp3,exp4,exp5,exp6,gatt_rev} Such studies are
 important both for basic scientific reasons and for possible
 technological applications.
 \cite{gatt_rev,qcomp}  Among the smaller SMM's are dimers of ${\rm V}^{4+}$ ($S=1/2$)
 and of ${\rm Fe}^{3+}$ ($S=5/2$), \cite{V2,Fe2} a nearly equilateral
 array of three ${\rm V}^{4+}$ spins, \cite{V3} and an isosceles
 triangle of ${\rm Gd}^{3+}$ ($S=7/2$) spins  (Gd3).\cite{Gd3} Non-metallic
 variations of 9$L$-BaRuO$_3$  with
 three  Ru$^{+4}$ ($S=1$) ions (Ru3) 
 might  behave as three-spin chain
 SMM's. \cite{9layer,pk}

  Each SMM consists of a small number of paramagnetic ions 
surrounded by
 non-magnetic chemical ligand groups, and is large enough that the
 magnetic interactions between the  ions in  different SMM's within a
crystal are negligible.
  Hence, measurements performed on
 macroscopic samples  just probe  the magnetic 
interactions within the individual SMM's.
 Large single crystals with long-range 
 SMM packing order are suitable for inelastic neutron scattering experiments.
Measurements at the appropriate deviations from the Bragg wave vectors for 
the particular SMM crystal structure  probe 
 statistical ensembles of the
Fourier transforms of the two-spin time correlation functions
involving the spins in each SMM.

 The magnetic interactions between the ions in these SMM's can
 often be described by the isotropic Heisenberg model. One expects that
 for ions such as Gd$^{3+}$, with $S=7/2$, the classical version of the
 Heisenberg model  captures the essential 
features of the dynamics at
 not too low temperatures $T$.
For a dimer,  comparisons of the  classical and $S=1/2, 5/2$ quantum
behaviors of the dynamics  
 supported this expectation.
 \cite{trans2,efremov-klemm} For the equilateral triangle, such
 comparisons were only made for $T\rightarrow\infty$. \cite{trans3}
 In addition, exact and numerical results
 were presented for the classical dimer \cite{muell-dim,muell-jphys,lubanAJP},
  the four-spin ring  \cite{4sp}, and the 
 $N$-spin equivalent neighbor
 model, which includes the equilateral triangle. \cite{EqNeigh}

For three spins on an equilateral triangle,
 with equal Heisenberg exchange constants $J_1$,  the spin sites 
 on each triangle  are  translationally
 invariant. 
Spins on an isosceles triangle have one additional exchange
 constant $J_2\ne J_1$, and this
 simple translational invariance is absent.  
Dynamical measurements on Gd3 and 
 modified compositions of Ru3 could help 
to uncover the effects
 of this lack of translational invariance inside each SMM, and  
  might aid in our understanding of  more 
 complicated  systems with multiple magnetic interactions.

At low $T$, the $N$ spins in the  equivalent neighbor model oscillate in a
single  mode. \cite{EqNeigh} 
Here we show that the absence of translational symmetry inside each 
isosceles triangle
 generally introduces 
 three additional low-$T$  modes  
 at tunable frequencies  depending
 upon $\gamma=J_2/J_1$.  For a semi-infinite
 range of $\gamma$ values,  
one of these
additional modes is a
 central peak.  In
 addition, the dynamics of the spins on the endpoints  of the 
three-spin chain with $J_2=0$ are
 qualitatively different from the  dynamics of the spin  at the chain center. 
The mode tuning parameter $\gamma$ makes the  dynamical behavior of the
 spins on an isosceles triangle  remarkably different from that present in 
any SMM system
 studied previously, including the four-spin ring. \cite{4sp,EqNeigh} 

 The structure of the paper is as follows: In Sec.\ \ref{Sec:Model}
 we define the model, calculate its partition function, and
 present the exact time evolution  of the spin vectors.
 In Secs.\ \ref{Sec:Correl} and  \ref{Sec:FT}, we discuss the time dependence
 and Fourier transform, respectively,  of each 
 spin-spin correlation functions.
  We discuss our results
  in Sec.\ \ref{Sec:Conclusions}. 

%
%

\section{The Model and the Spin Dynamics}      \label{Sec:Model}

 We study the  Hamiltonian describing three spins
 ${\bf S}_i$ of unit magnitude, $S_i=|{\bf S}_{i}|=1$, on a  triangle with two
 classical Heisenberg exchange couplings,
\begin{equation}        \label{H}
        H = -J_{1} ( {\bf S}_{1} \cdot {\bf S}_{2} +
            {\bf S}_{2} \cdot {\bf S}_{3}) -
            J_{2} {\bf S}_{1} \cdot {\bf S}_{3}.
\end{equation}
 The cases $J_1=J_2$ and $J_{1} \neq J_{2}\ne0$ describe equilateral and isosceles
 triangles, respectively. The case
 $J_{2}=0$ describes a three-spin  chain with
 free boundary conditions. 
Equation\ (\ref{H}) can be realized if the ring contains either
 two different lattice constants, as in Gd3, or ions with two different
 spin values. 

 We rewrite Eq.\ (\ref{H}) in terms of the total spin
 ${\bf S}={\bf S}_{2} + {\bf S}_{13}$, with ${\bf S}_{13} =
 {\bf S}_{1} + {\bf S}_{3}$, and obtain, up to a constant,
\begin{equation}        \label{newH}
        H = - \frac{J_{1}}{2} {\bf S}^{2} - \frac{J_{2}-J_{1}}{2}
            {\bf S}_{13}^{2}.
\end{equation}
 The partition function $Z=\int (\prod_{i=1}^{3} d\Omega_{i}/
 4\pi) e^{-\beta H}$ is
\begin{eqnarray}
        Z &=& \int_{0}^{2} dx \int_{|x-1|}^{x+1} ds \, s \, e^{-\beta H}
          \nonumber \\
          &=& \frac{e^{\alpha}}{\alpha} \int_{0}^{2} dx e^{\alpha \gamma x^2}
            \sinh (2\alpha x),
        \label{Z}
\end{eqnarray}
 where $\beta=(k_{B}T)^{-1}$, $d\Omega_{i}$ is the solid angle element
 for the $i$-th spin, $\alpha = \beta J_{1}/2$,
 $\gamma = J_{2}/J_{1}$, $x=S_{13}$, and $s=S$.

 To calculate the time-dependent correlation functions
 $\langle {\bf S}_{i}(t)\cdot {\bf S}_{j}(0) \rangle $, we first solve
 the classical Heisenberg equations of motion for the quantities
 ${\bf S}_{2}(t)$ and ${\bf S}_{13}(t)$,
\begin{equation}        \label{eqm}
        \dot{{\bf S}}_{2,13} = J_{1} {\bf S}_{2,13} \times {\bf S},
\end{equation}
 leading to

\begin{equation}        \label{S2.13}
        {\bf S}_{2,13}(t) = C_{2,13} \hat{{\bf s}} + A_{2,13} \left[
          \cos (J_{1}st) \hat{{\bf x}} - \sin (J_{1}st) \hat{{\bf y}} \right],
\end{equation}
 where $\hat{\bf s}||{\bf S}$, $\hat{{\bf s}}=\hat{{\bf x}} \times \hat{{\bf y}}$, 
 $A_{2}=-A_{13}$, $C_{2}+C_{13}=s$,
  $C_{13} = (s^2+x^2-1)/(2s)$, and $A_{13}^{2} = x^{2}-C_{13}^2$.

 The time dependence of ${\bf S}_{1}$ (or ${\bf S}_{3}$) is obtained from
\begin{equation}
        \dot{{\bf S}}_{1} = (J_{2}-J_{1}) {\bf S}_{1} \times {\bf S}_{13}
          + J_{1} {\bf S}_{1} \times {\bf S} ,
\end{equation}
 leading to %
\begin{eqnarray}
        S_{1s}(t)& =& S_{1s0}+ \Delta S_{1s0} \cos[(J_{1}-J_{2})xt-\phi_{0}],
          \label{S1s}\\
        S_{1\pm}(t) &=& \frac{A_{13}S_{1s0}}{C_{13}} \exp (\mp isJ_{1}t)
          \nonumber \\
        & & \hspace{-34pt} +
          \frac{A_{13} \Delta S_{1s0}}{2(C_{13}+x)} \exp \left\{
          \mp i[(J_{1}s+(J_{1}-J_{2})x]t \pm i\phi_{0} \right\} \nonumber \\
        & & \hspace{-34pt} + \frac{A_{13} \Delta S_{1s0}}{2(C_{13}-x)}
          \exp \left\{ \mp i[(J_{1}s-(J_{1}-J_{2})x]t \mp i\phi_{0} \right\},
          \label{S1pm}
\end{eqnarray}
 where $S_{1\pm}=S_{1x}\pm i S_{1y}$, $S_{1s0} = C_{13}/2$, 
 $\Delta S_{1s0} = A_{13}(1- x^2/4)^{1/2}/x$, and  $\phi_{0}$  is an arbitrary
 angle describing the initial spin configuration. \cite{4sp}

%
%

\section{Time Correlation Functions}      \label{Sec:Correl}

 We study  the two-spin time correlation functions
\begin{equation}        \label{Cij}
        {\cal C}_{ij}(t) = \langle {\bf S}_{i}(t) \cdot {\bf S}_{j}(0)
          \rangle ,
\end{equation}
 where  $\langle \ldots \rangle=\int_0^{2\pi}d\phi_0\int_0^2dx
\int_{|x-1|}^{x+1}sds e^{-\beta H}\ldots/(2\pi Z)$. Since ${\cal C}_{11}={\cal C}_{33}$ and
 ${\cal C}_{12}={\cal C}_{23}$, only four ${\cal C}_{ij}$ are
  independent.  These are constrained by the sum rule
\begin{equation}        \label{sumrule}
        \langle s^{2} \rangle = {\cal C}_{22}(t) + 2 {\cal C}_{11}(t) +
          4 {\cal C}_{12}(t) + 2 {\cal C}_{13}(t).
\end{equation}
After averaging over $\phi_0$,
 the exact  ${\cal C}_{ij}(t)$ satisfy
\begin{eqnarray}
        {\cal C}_{11}(t) &=& I_{0} + I_{1}(t) + I_{2}(t) + I_{3}(t),
                \label{C11-I}\\
        {\cal C}_{13}(t) &=& I_{0} + I_{1}(t) - I_{2}(t) - I_{3}(t),
                \label{C13-I}\\
        {\cal C}_{22}(t) &=& 4 \left[I_{0} + I_{1}(t) \right] +
          2 \langle C_{2}s \rangle - \langle s^{2} \rangle,
                \label{C22-I}\\
        {\cal C}_{12}(t) &=& -2 \left[I_{0} + I_{1}(t) \right] +
          \frac{1}{2} \left[ \langle s^{2} \rangle - \langle C_{2}s \rangle
          \right],
                \label{C12-I}
\end{eqnarray}
where by setting $t^*=J_1t$, we have
\begin{eqnarray}
        I_{0} &=& \frac{1}{4} \langle C_{13}^{2} \rangle, \label{i0} \\
        I_{1}(t) &=& \frac{1}{4} \langle A_{13}^{2} \cos (st^{*}) \rangle ,
          \label{i1} \\
        I_{2}(t) &=& \frac{1}{2} \langle \frac{A_{13}^{2}}{x^{2}}
                \left( 1 - \frac{x^{2}}{4} \right) \cos \left[
                (1-\gamma)xt^{*} \right] \rangle , 
          \label{i2}\\
        I_{3}(t) &=& \langle \frac{(1-x^{2}/4)}{4x^2}
          \Bigl( (C_{13}-x)^{2} \cos \{[s+(1-\gamma)x]t^{*}\} 
          \nonumber \\
        & & +  (C_{13}+x)^{2} \cos \{[s-(1-\gamma)x]t^{*}\} \Bigr)
          \rangle .
          \label{i3}
\end{eqnarray}

 As $T\rightarrow \infty$, the remaining double integrals in Eqs.\
 (\ref{i0})--(\ref{i3}) can be reduced to single integrals, 
as shown in the Appendix.  Note that $I_2(t)$ is a constant for $\gamma=1$, but
 not for $\gamma\ne1$, leading to dramatic differences between the dynamics
 of spins on  equilateral and isosceles  triangles, respectively. 
 As $t \rightarrow \infty$, the\ time-dependent
 integrands oscillate wildly and their contributions vanish.
 \cite{note-Riemann}
  From Eqs.\ (\ref{C22-I}) and (\ref{i1}), $\lim_{T,t\rightarrow \infty}
{\cal C}_{22}$ for $\gamma\ne1$ is
 identical to its value for $\gamma=1$.
  Only at finite $T$ does $J_{2}$
 influence ${\cal C}_{22}(t)$. Hence, for $t^*\gg1$ and $T\rightarrow\infty$,
  \cite{4sp,delta3}
\begin{eqnarray}        \label{C22longt}
        \lim_{{T \rightarrow \infty}\atop{t^{*} \gg 1}} {\cal C}_{22}(t)
 &\sim &
          \frac{1}{3} + 2 \, \delta_{3} - \frac{\sin (t^{*}) +
          \sin(3t^{*})}{(t^{*})^{3}},\\
\delta_3& =& {9\over{80}}\ln3-{1\over{20}}\approx 0.07359.\label{d3}
\end{eqnarray}
 From Eqs.\ (\ref{C22-I})--(\ref{C12-I}), it then follows that
\begin{equation}
        \lim_{{T \rightarrow \infty}\atop{t^{*} \gg 1}} {\cal C}_{12}(t)
          \sim \frac{1}{3} - \delta_{3} + \frac{\sin (t^{*}) +
          \sin(3t^{*})}{2(t^{*})^{3}}.
\end{equation}

 The surprise comes from the behaviors of
 ${\cal C}_{11}$ and  ${\cal C}_{13}$. The long-time 
 approaches to the various asymptotic limits are characterized by three 
different powers, corresponding
 to the equilateral, the isosceles, and the chain cases, respectively.
 For $\gamma =1$, 
 ${\cal C}_{11}^{\gamma =1}(t) = {\cal C}_{22}(t)$, which
 approach their mutual limit as $t^{-3}$ as $T \rightarrow \infty$.

 For $\gamma =0$, ${\cal C}_{11}(t)$ is dominated by $I_{3}(t)$, yielding
\begin{equation}        \label{C11longtchain}
        \lim_{{T \rightarrow \infty}\atop{t^{*} \gg 1}}
          {\cal C}_{11}^{\gamma =0}(t) \sim
          \frac{1}{3} + \frac{\delta_{3}}{2}
          + \frac{1}{2t^{*}} \sin (t^{*}),
\end{equation}
which approaches its different asymptotic limit dramatically slower.

 For $\gamma\ne 0, 1$, $I_{2}(t)$ and $I_{3}(t)$ decay
 as $t^{-2}$ and oscillate with different frequencies.
 The evaluation of $I_{3}(t)$ is quite involved. By
 calculating its Fourier transform 
 and then inverting it through integration by parts, we find,
\begin{eqnarray}
        \lim_{{T \rightarrow \infty}\atop{|\gamma|
            t^{*},|1-\gamma|t^{*} \gg 1}}
          {\cal C}_{11}^{\gamma \neq 0,1}(t)
          & \sim &  \frac{1}{3} + \frac{\delta_{3}}{2} \nonumber \\
        & & \hspace{-62pt} + \frac{ \left\{ a_{1}
          + a_{2} \cos [2(1-\gamma)t^{*}] 
          + a_{3} \cos (t^{*}) \right\} }{ (t^{*})^{2} },
        \label{C11longt}
\end{eqnarray}
 where (see Appendix)
\begin{eqnarray}
        a_{1} &=& -\frac{1}{6(1-\gamma)^{2}} ,
          \label{a1} \\
        a_{2} &=& -\frac{1}{(1-\gamma)^{2}} \left(
          \frac{5}{32} - \frac{9}{128} \ln 3 \right)
          \approx -\frac{0.079004}{(1-\gamma)^{2}} ,
          \label{a2} \\
        a_{3} &=& - \frac{\gamma -1}{2\gamma} +
          \frac{\gamma -2}{4} \ln \left| \frac{\gamma}{\gamma -2}
          \right| . \label{a3}
\end{eqnarray}

 Note that one cannot take either of the limits $\gamma \rightarrow 1$
 or $\gamma \rightarrow 0$ directly in Eq.\ (\ref{C11longt}), since
 the expansion is valid only when both $|\gamma| t^{*}\gg 1$ and
 $|1-\gamma|t^{*} \gg 1$ are satisfied.

\begin{figure}
 \epsfxsize=8.5cm
 \centerline{\epsffile{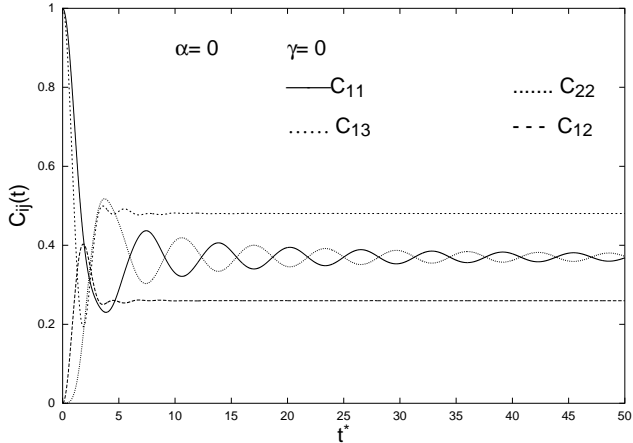}}
 \vspace{0.3cm}
 \caption{Plots of ${\cal C}_{11}(t)$ (solid), ${\cal C}_{12}(t)$ (dashed),
  ${\cal C}_{13}(t)$ (dense-dotted, ${\cal C}_{13}(0)=0$),
  ${\cal C}_{22}(t)$ (sparse-dotted, ${\cal C}_{22}(0)=1$)
  versus $t^{*}=|J_{1}|t$ as $T\rightarrow \infty$ ($\alpha=0$)
  for the chain ($\gamma=0$).}
 \label{akfig1}
\end{figure}

 In Fig.\ \ref{akfig1}, we plot the
 ${\cal C}_{ij}(t)$ as
 $T\rightarrow\infty$ for $\gamma=0$.  Since
  ${\cal C}_{22}(t)$
 and ${\cal C}_{12}(t)$ as $T\rightarrow\infty$ are independent of
 $\gamma$, these curves are respectively identical to those for ${\cal C}_{11}(t)$ and ${\cal
 C}_{12}(t)$  obtained for
 $T\rightarrow\infty$ 
 in the equilateral triangle. These functions
 each approach their asymptotic limits as
 $t^{-3}$.  On the other hand, for $\gamma=0$, 
  ${\cal C}_{11}(t)$ and ${\cal C}_{13}(t)$ oscillate
 about each other for a long time, approaching their mutual asymptotic
 limit as $t^{-1}$.

\begin{figure}
 \epsfxsize=8.5cm
 \centerline{\epsffile{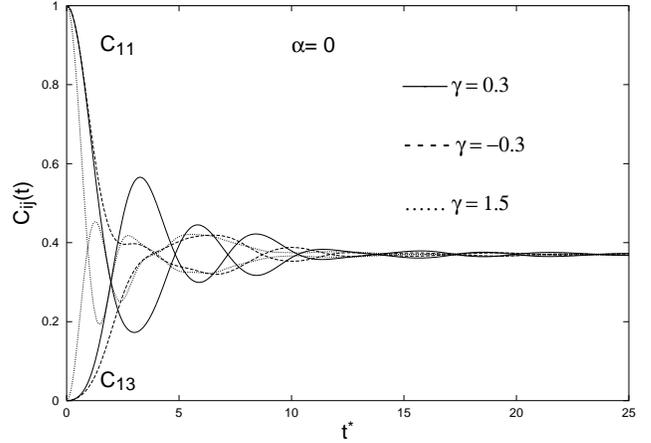}}
 \vspace{0.3cm}
 \caption{Plots of ${\cal C}_{11}(t)$, ${\cal C}_{13}(t)$ versus
  $t^{*}=|J_{1}|t$ for $T\rightarrow \infty$ ($\alpha$=0),
  $\gamma=0.3$ (solid), -0.3 (dashed), and 1.5 (dotted).}
 \label{akfig2}
\end{figure}

 In Fig.\ \ref{akfig2} we compare  ${\cal C}_{11}(t)$ and
 ${\cal C}_{13}(t)$ as $T\rightarrow\infty$ for $\gamma=\pm 0.3$ and 
 1.5. For $\gamma=\pm 0.3$, the leading term in the approach to the
 asymptotic limit arises from $a_3$ in Eq.\ (\ref{C11longt}).  For
 $\gamma=1.5$,  $2|1-\gamma|=1$, so there is asymptotically only one 
frequency,  equal to
 that for the chain shown in Fig.\ 1,
 but the asymptotic limit  is approached faster.

 At finite $T$, the physics of the  model is influenced
 not only by  $\gamma$, but also by the sign of $J_{1}$. We henceforth
 refer to the   $J_{1}>0$ and $J_1<0$ cases
 as ferromagnetic (FM) and  antiferromagnetic
 (AFM), respectively.

\begin{figure}
 \epsfxsize=8.5cm
 \centerline{\epsffile{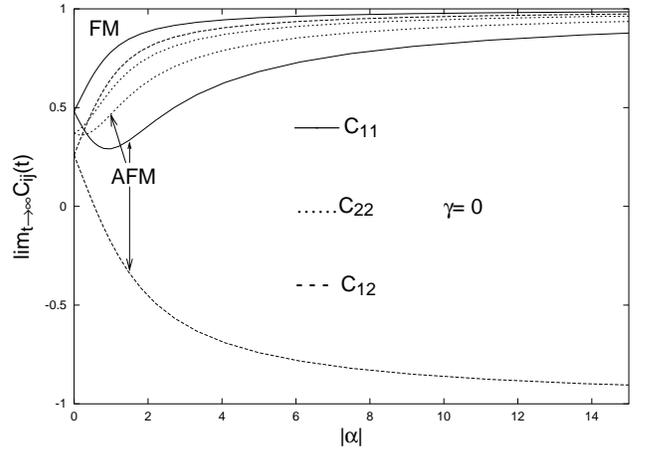}}
 \vspace{0.3cm}
 \caption{Plots of $\lim_{t\rightarrow \infty} {\cal C}_{ij}(t)$ for the
  chain ($\gamma$=0) versus $|\alpha|=\beta |J_{1}|/2$,
  for $\alpha >0$ (FM) and $\alpha <0$ (AFM).}
 \label{akfig3}
\end{figure}

 We obtain the ${\cal C}_{ij}(t)$ at finite
 $T$ by direct numerical evaluation of
 the double integrals  in Eqs.\ (\ref{C11-I})-(\ref{C12-I}).
 As an example, in 
  Fig.\ \ref{akfig3}, we plot the
 $\lim_{t\rightarrow \infty} {\cal C}_{ij}(t)$ for 
 $\gamma=0$ as functions of
 $|\alpha| = \beta |J_{1}|/2$. The FM and AFM cases are distinguished
 by arrows. We recall that $\lim_{t\rightarrow \infty} {\cal C}_{13}(t)=
 \lim_{t\rightarrow \infty} {\cal C}_{11}(t)$ for all $\gamma \neq 1$.
 The spins are intrinsically unfrustrated. At low $T$,  the FM 
 $\lim_{t\rightarrow \infty} {\cal C}_{ij}(t)\sim 1$, 
 while  the AFM $\lim_{t\rightarrow \infty}
 {\cal C}_{12}(t)\sim -1$. In the AFM case, $\lim_{t\rightarrow
   \infty} {\cal C}_{11}(t)$ has a minimum value of $\sim 0.29$ at
 $\alpha \sim -0.9$.

%
%

\section{Fourier Transforms}    \label{Sec:FT}

 The dimensionless Fourier transform (FT) of each deviation
 $\delta {\cal C}_{ij}(t) \equiv {\cal C}_{ij}(t) - 
 \lim_{t \rightarrow \infty} {\cal C}_{ij}(t)$, \cite{4sp} is
\begin{equation}
        \delta\tilde {\cal C}_{ij}(\omega) =\frac{|J_{1}|}{\pi}
          \int_{-\infty}^{+\infty}
          dt e^{i\omega t} \delta {\cal C}_{ij}(t).\label{ft}
\end{equation}
 Since causality requires $\delta \tilde{{\cal C}}_{ij}(\omega) =
 \delta \tilde{{\cal C}}_{ij}(-\omega)$, we consider only
 positive values of $\omega/J_1$.  The Fourier integral is in
 principle elementary, since the time dependence is contained in
 simple trigonometric factors, yielding combinations of
 $\delta$-functions that allow us to perform one of the integrations.
 We thus obtain single integral representations for the FT's at any
 $T$, which can be evaluated with high precision. 
 Since each integration domain is restricted to a finite region in
 the $(s,x)$ plane, the $\delta$-functions only  contribute to the integrals
 if $\omega$ falls within  specific ranges.
 Determining the allowed ranges of $\omega$ values is
 tedious.
Some of the details of this calculation  are presented in
 the Appendix. 

\subsection{Infinite temperature}

\begin{figure}
 \epsfxsize=8.5cm
 \centerline{\epsffile{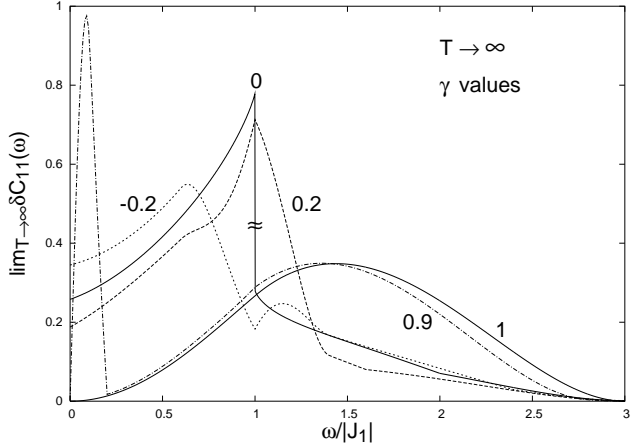}}
 \vspace{0.3cm}
 \caption{Plots of  $\delta
  \tilde{\cal C}_{11}(\omega)$ as functions of $\omega/|J_1|$ as
  $T\rightarrow\infty$, for  $\gamma=1,0$ (solid), 0.2 (dashed), -0.2
 (dotted), and 0.9 (dash-dotted).}
 \label{akfig4}
\end{figure}

 In Fig.\ \ref{akfig4} we plot $\delta \tilde{\cal C}_{11}(\omega)$
 as $T\rightarrow\infty$ for some $|\gamma| \le 1$. The solid curves for 
$\gamma=1,0$
 correspond to the equilateral triangle and the chain.  For the equilateral
 triangle, $\delta\tilde{\cal
 C}_{11}(\omega)$ vanishes as $\omega\rightarrow0,3$, and exhibits a single
 smooth peak at $\omega/|J_1|\approx 1.4385.$ Although difficult to discern in this
 figure, the curve contains discontinuous second derivatives at
 $\omega/|J_1|=1,3$ responsible for
 the $t^{-3}$ long-time, $T\rightarrow\infty$ decay of 
$\delta{\cal C}_{11}(t)$. \cite{4sp,EqNeigh} 

 As $T\rightarrow\infty$, although $\delta\tilde{\cal C}_{22}(\omega)$ for the
 chain coincides
 with the equilateral triangle $\tilde\delta{\cal C}_{11}(\omega)$ curve,
 the chain $\delta\tilde{\cal C}_{11}(\omega)$ is dramatically different. In addition to the slope discontinuity at $\omega/|J_1|=2$,   
 the chain $\delta\tilde{\cal C}_{11}(\omega)$  is
 {\it discontinuous} at $\omega/|J_1|=1$, as indicated by the symbol
 $\approx$.  This discontinuity 
 is responsible for the $1/t$ long-time, $T\rightarrow\infty$ decay of 
$\delta{\cal C}_{11}(t)$. 

The isosceles triangle cases are also very interesting, as they do not just
interpolate between the equilateral triangle and the chain.   For
 $\gamma=\pm0.2$, the discontinuities in  the  slope of $\delta\tilde{\cal
C}_{11}(\omega)$ at 
$\omega /|J_{1}|=1,1.4,$ and 1.6
  are also  discernible.  As $\gamma \rightarrow 0$, the
 slope discontinuities at $\omega/|J_1|=1$ 
 approach $\pm \infty$  as expressed in
 Eq.\ (\ref{a3}). In addition, there is a  large discontinuity in the slope of
 the curve for $\gamma=0.9$ 
at $\omega/J_1=0.2$, and a resulting large peak at $\omega/|J_1|\approx
0.088$, arising from
 $\tilde{I}_{2}(\omega)$. This slope discontinuity is
 responsible for the first two terms  $\propto t^{-2}$ in Eq.\
 (\ref{C11longt}). 

\subsection{Low temperature}

 The  $\delta\tilde{\cal C}_{ij}(\omega)$ at finite $T$ were obtained by numerical evaluation of the
 single integrals defined in the Appendix. As $T\rightarrow\infty$, we
 compared our numerical and analytic results. At finite $T$ we also checked
 the initial value sum rules by integrating the FT's in Eq. (\ref{ft}).

\subsubsection{Equilateral triangle}

 \begin{figure}
 \epsfxsize=8.5cm
 \centerline{\epsffile{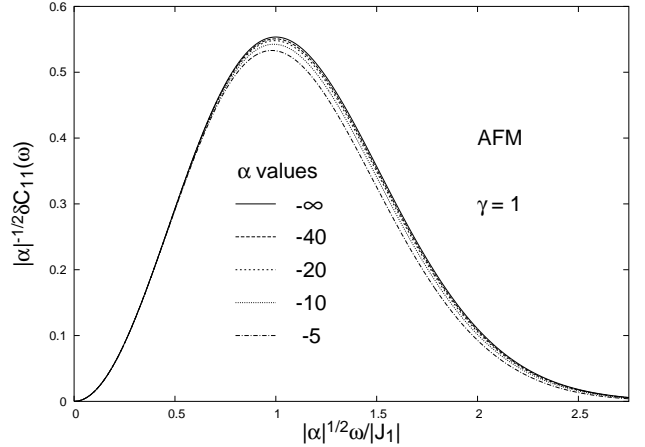}}
 \vspace{0.3cm}
 \caption{Plots of  $|\alpha|^{-1/2}\delta
  \tilde{\cal C}_{11}(\omega)$ versus $|\alpha|^{1/2}\omega /|J_{1}|$ for the
 AFM equilateral triangle at low $T$. }
 \label{akfig5}
\end{figure}

  We first discuss the  equilateral
 triangle, $\gamma=1$.  For the AFM case, As $T$ decreases, the single peak in 
$\delta\tilde{\cal
 C}_{11}(\omega)$ grows in amplitude and shifts to lower $\omega/|J_1|$.
 As shown for the equivalent neighbor model, \cite{EqNeigh}, this behavior can
 be quantified by plotting 
$|\alpha|^{-1/2}\delta\tilde{\cal C}_{11}(\omega)$
versus $\tilde{\omega}=|\alpha|^{1/2}\omega/|J_1|$.
 In Fig.\ \ref{akfig5},
   the
 $|\alpha|^{-1/2}\delta\tilde{\cal C}_{11}(\omega)$ curves approach the
 uniform AFM equivalent neighbor model form
 ${8\over{3\sqrt{\pi}}}\tilde{\omega}^2\exp(-\tilde{\omega}^2)$, shown as the
 solid curve, as
 $T\rightarrow0$. \cite{EqNeigh} This frustrated behavior
 results in a scaling of the
 time, as ${\cal C}(t)$ approaches a uniform function of $(JT)^{1/2}t$ as
 $T\rightarrow0$.

\begin{figure}
 \epsfxsize=8.5cm
 \centerline{\epsffile{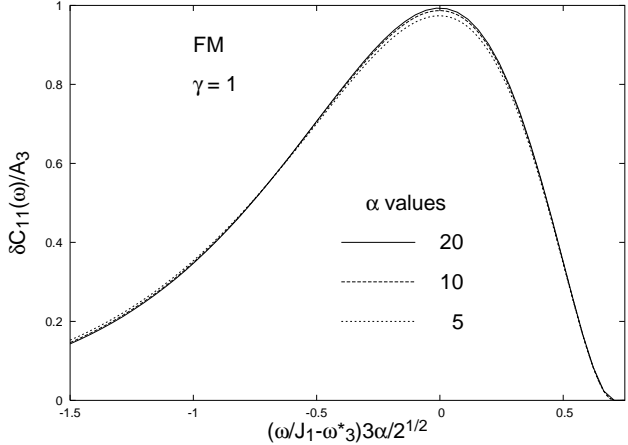}}
 \vspace{0.3cm}
 \caption{Plots of  $\delta
  \tilde{\cal C}_{11}(\omega)/A_3$ versus $(\omega /J_{1}-\omega_3^*)3\alpha/2^{1/2}$ for the
 FM equilateral triangle at low $T$, where $A_3=8/(3e^2)$ and $\omega_3^*=3-1/(3|\alpha|)$. }
 \label{akfig6}
\end{figure}

For the FM equilateral triangle, as $T$ decreases, the peak in
$\delta\tilde{\cal C}_{11}(\omega)$ shifts asymptotically to $3J_1$, as all of
the spins oscillate together.  In the equivalent neighbor model,
as $T\rightarrow0$, the three-spin $\gamma=1$ FM curves approach a uniform
function of $(\omega/J_1-\omega^*_3)\alpha$, where
$\omega^*_3=3-1/(3|\alpha|)$, as shown in Fig.\
\ref{akfig6}. \cite{EqNeigh} That is, as $T\rightarrow0$, the peak amplitude tends to a
constant value, but its maximum position approaches $3J_1$ linearly in $T$.   

\subsubsection{Three-spin chain}

\begin{figure}
 \epsfxsize=8.5cm
 \centerline{\epsffile{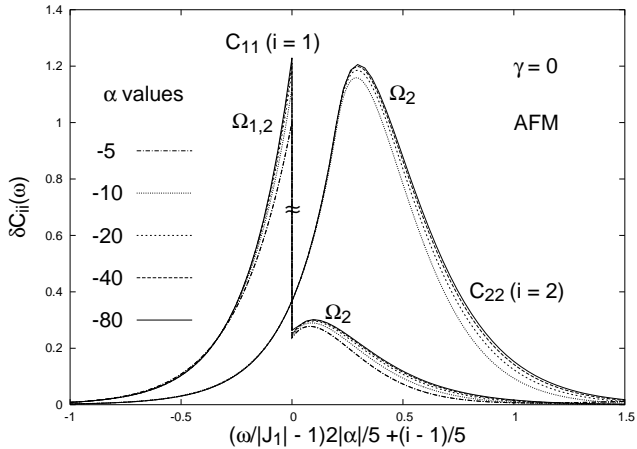}}
 \vspace{0.3cm}
 \caption{Plots of $\delta\tilde{\cal C}_{11}(\omega)$ versus
 $(\omega/|J_1|-1)2|\alpha|/5$ and $\delta\tilde{\cal C}_{22}(\omega)$ versus
 $(\omega/|J_1|-1)2|\alpha|/5$ +0.2, for the AFM  chain at the low-$T$ values
 $-\alpha=5, 10, 20,
 40, 80$.  $\tilde{\cal C}_{11}(\omega)$ is discontinuous at
 $\omega/|J_1|=1$, indicated by the $\approx$.}
 \label{akfig7}
\end{figure}

Strikingly different behavior is obtained for the low-$T$ AFM chain. 
In this case, instead of the single AFM mode present in the equilateral
triangle, there are {\it four} modes, the frequencies $\Omega_i$ of which approach
$\Omega_i/|J_1|=1, 2, 3$ as $T\rightarrow0$, as the frequencies of the
strongest modes, 
$\Omega_{1}, \Omega_2$, become degenerate.      
In Fig.\ \ref{akfig7}, we plotted $\delta\tilde{\cal C}_{11}(\omega)$ and
$\delta\tilde{\cal C}_{22}(\omega)$ for
these two modes, versus $(\omega/|J_1|-1)2|\alpha|/5$, at various low-$T$
$\alpha$ values.  Curves for $\delta\tilde{\cal C}_{22}(\omega)$ are shifted to
the right by 0.2 for clarity.  We note that 
$\delta\tilde{\cal C}_{11}(\omega)$ 
for the $\Omega_2$
mode has a shallow maximum at a frequency
which approaches $|J_1|$ from above as $T\rightarrow0$.  For the $\Omega_1$
mode, $\delta\tilde{\cal C}_{11}$ has a large discontinuity precisely at
$\omega/|J_1|=1$ for all $T$.  By contrast, the $\delta\tilde{\cal
C}_{22}(\omega)$ curves only exhibit the $\Omega_2$ mode, and are smooth
and rather symmetric about their maxima.  Both $\delta\tilde{\cal
C}_{ii}(\omega)$ for these modes approach a
uniform scaling function of $|\alpha|(\omega/|J_1|-1)$ as $T\rightarrow0$.
We remark that the low-$T$ scaling behaviors of the dominant modes in the 
AFM chain are similar to
that of the FM equilateral triangle, since neither of these mode energies
approaches 0 as $T\rightarrow0$.

In addition, we found that the weaker $\Omega_3$ and $\Omega_4$ modes, for
which $\delta\tilde{\cal C}_{11}(\omega)$ is peaked at $\omega/|J_1|\approx
3,2$, respectively,  both approach
uniform low-$T$ scaling functions. For the $\Omega_3$ mode, $|\alpha|\delta\tilde{\cal
C}_{11}(\omega)$ approaches a uniform scaling function of
$|\alpha|(\omega/|J_1|-3)$, and for the $\Omega_4$ mode,
$\alpha^2\delta\tilde{\cal C}_{11}(\omega)$ approaches a uniform scaling
function of $|\alpha|(\omega/|J_1|-2)$ as $T\rightarrow0$.  Hence, as
$T\rightarrow0$, the amplitudes of these modes vanish as $T$ and $T^2$,
respectively.

\begin{figure}
 \epsfxsize=8.5cm
 \centerline{\epsffile{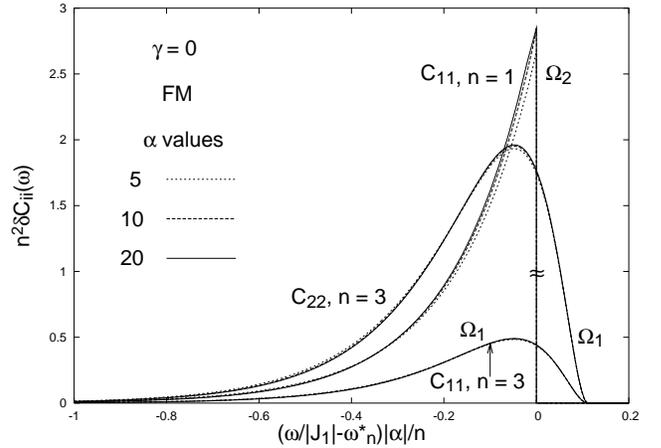}}
 \vspace{0.3cm}
 \caption{Plots of $n^2\delta\tilde{\cal C}_{11}(\omega)$  and 
$n^2\delta\tilde{\cal C}_{22}(\omega)$ versus
 $(\omega/|J_1|-\omega_n^*)|\alpha|/n$ for the $\Omega_1$ and $\Omega_2$ modes
 of the FM  chain at the low-$T$ values
 $\alpha=5, 10, 20$, where $\omega_n^*=n-(n-1)/(2n|\alpha|)$.  $\tilde{\cal
 C}_{11}(\omega)$ for the $\Omega_2$ mode is discontinuous at
 $\omega/|J_1|=1$, as indicated by the $\approx$.}
 \label{akfig8}
\end{figure}

For the FM chain at low $T$, there are  four non-degenerate modes for
which $\delta\tilde{\cal C}_{11}(\omega)$ is peaked at $\Omega_i/|J_1|\approx 1, 2,
3, 5$.  At low $T$, the two largest intensity modes $\Omega_2$ and
$\Omega_1$, which are peaked at $\omega/J_1\approx 1,3$, respectively, are 
pictured for $\alpha=5,10,20$ in Fig.\ \ref{akfig8}.  At low $T$,
$\delta\tilde{\cal C}_{11}$ for the $\Omega_2$ mode is very large for $\omega/|J_1|<1$, but drops
{\it discontinuously to zero} at $\omega/|J_1|=1$, as shown in 
Fig.\ \ref{akfig8}.  For the next largest intensity mode, $\Omega_1$, both
$\delta\tilde{\cal C}_{11}(\omega)$ and the larger $\delta\tilde{\cal
C}_{22}(\omega)$ exhibit peak positions that approach $\omega/|J_1|=3$ as
$T\rightarrow0$, as shown in Fig.\ \ref{akfig8}. We note that we have
multiplied the intensities of these curves by a factor of 9, and squeezed the
scaling variable by a factor of 3, relative to those of $\delta\tilde{\cal
C}_{11}$ for the $\Omega_2$ mode, in order to fit all three sets of curves in
the same figure. All of the curves shown in
Fig.\ \ref{akfig8} scale as for the FM equilateral triangle. As for the AFM
chain, at low $T$, $\alpha\delta\tilde{\cal C}_{11}(\omega)$ for the
$\Omega_4$ mode approaches a uniform
function of $(\omega/|J_1|-2)|\alpha|$, and $\alpha^2\delta\tilde{\cal
C}_{11}(\omega)$ for the weakest  mode $\Omega_3$ approaches a uniform function
of $(\omega/|J_1|-5)|\alpha|$. 

\subsubsection{General isosceles triangle}

We now consider the more general isosceles triangle cases.  We 
first discuss the AFM cases, $J_1<0$.
As for the chain, for $\gamma\ne1$, as $T$ is lowered,
 $\delta\tilde{\cal C}_{11}(\omega)$
 generally develops into three or four  peaks, which
 become progressively sharper. When only three modes are present, one of them
is a central peak at $\omega=0$.  For non-vanishing frequencies,
 these  modes  are  magnons. As for the three-spin chain, their relative
 intensities at finite $T$ are very different, as the two largest are typically a few orders
 of magnitude larger than the third, which is 
  a few orders of magnitude larger than the fourth. 

\begin{figure}[t]
 \epsfxsize=8.5cm
 \centerline{\epsffile{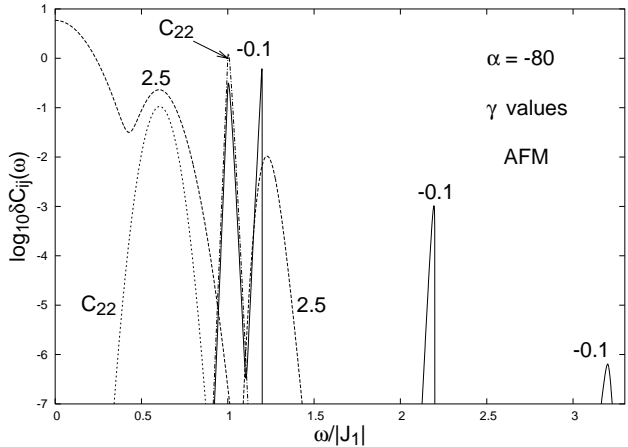}}
 \vspace{0.3cm}
 \caption{Plots of the AFM
  ${\rm log}_{10}[\delta\tilde{\cal C}_{ij}(\omega)]$
  versus $\omega/|J_1|$, at low $T$ ($\alpha=-80$) for  $\gamma=-0.1,
 2.5$.  The solid (dashed) curves correspond to $\delta\tilde{\cal
 C}_{11}(\omega)$ for $\gamma=-0.1$ (2.5).  The dash-dotted (dotted) curves
 are the corresponding $\delta\tilde{\cal C}_{22}(\omega)$ curves.}
 \label{akfig9}
\end{figure}

 To illustrate the types of low-$T$ AFM behavior, in Fig.\ \ref{akfig9},  
we plot
 $\log_{10}[\delta\tilde{\cal C}_{11}(\omega)]$ and
 $\log_{10}[\delta\tilde{\cal C}_{22}(\omega)]$ versus $\omega/|J_1|$ at the
 AFM low-$T$ value
 $\alpha=-80$ for $\gamma=-0.1, 2.5$.  For $\gamma=-0.1$, $\delta\tilde{\cal
C}_{11}(\omega)$ exhibits four sharp peaks at $\omega/|J_1|\approx$ 1, 1.2,
2.2, and 3.2.  For $\gamma=2.5$, there are three broad peaks in  $\delta\tilde{\cal
C}_{11}(\omega)$, one of which is a central peak at $\omega=0$, and the others
are centered at $\omega/|J_1|\approx$ 0.6 and 1.2.  In each case, 
 $\delta\tilde{\cal C}_{22}(\omega)$ is a single peak at
one of the larger
non-vanishing 
$\delta\tilde{\cal C}_{11}(\omega)$ peak positions.

\begin{figure}
 \epsfxsize=8.5cm
 \centerline{\epsffile{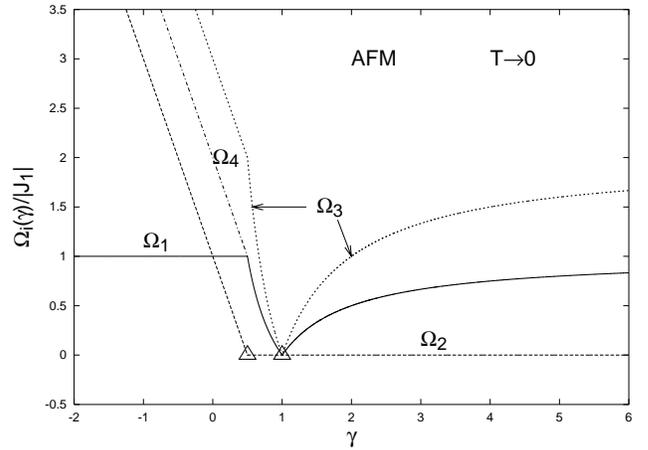}}
 \vspace{0.3cm}
 \caption{Plots of the low-$T$ AFM mode frequencies $\Omega_i(\gamma)/|J_{1}|$
versus $\gamma$. The
  triangles  indicate low-$T$ scaling.}
 \label{akfig10}
\end{figure}

 After a careful analysis of many lower-$T$ results, we 
 established simple formulas relating the frequencies of the modes to
 $\gamma$. $\delta\tilde{\cal C}_{11}(\omega)$ contains four low-$T$ AFM mode
 frequencies, $\Omega_i(\gamma)$,  depicted in Fig.\ \ref{akfig10}, which
satisfy
\begin{eqnarray}
        \Omega_{1}(\gamma)/|J_{1}| &=& 1,~~{\rm for}~~ \gamma \leq 1/2 ,
           \nonumber \\
                   &=& \left| 1-1/ \gamma \right|, ~~
                       {\rm for}~~ \gamma \geq 1/2 ;
                \label{Om1AFM} \\
        \Omega_{2}(\gamma)/|J_{1}| &=& 1-2\gamma ,~~{\rm for}~~ \gamma
                \leq 1/2 , \nonumber \\
                   &=& 0,~~{\rm for}~~ \gamma \geq 1/2 ;
                \label{Om2AFM} \\
        \Omega_{3}(\gamma)/|J_{1}| &=& 3-2\gamma ,~~{\rm for}~~ \gamma
                \leq 1/2 , \nonumber \\
                   &=& 2 \left| 1-1/\gamma \right|, ~~
                       {\rm for}~~ \gamma > 1/2 ;
                \label{Om3AFM} \\
        \Omega_{4}(\gamma)/|J_{1}| &=& 2-2\gamma ,~~{\rm for}~~ \gamma
                \leq 1/2 , \nonumber \\
                   &=& \left| 1-1/ \gamma \right|, ~~
                       {\rm for}~~ \gamma \geq 1/2 .
                \label{Om4AFM}
\end{eqnarray}

$\delta \tilde{{\cal C}}_{22}(\omega)$ contains only
 the  low-$T$ FM mode with frequency $\Omega_1(\gamma)$.  We note that for $\gamma\ge1/2$, $\Omega_4$ and $\Omega_1$ are degenerate.
   We  analytically confirmed
 Eq.\ (\ref{Om1AFM})  by performing an
 asymptotic evaluation of the integral representation of 
$\delta\tilde{\cal C}_{22}(\omega)$, as shown 
  in the Appendix. 
 
\begin{figure}
 \epsfxsize=8.5cm
 \centerline{\epsffile{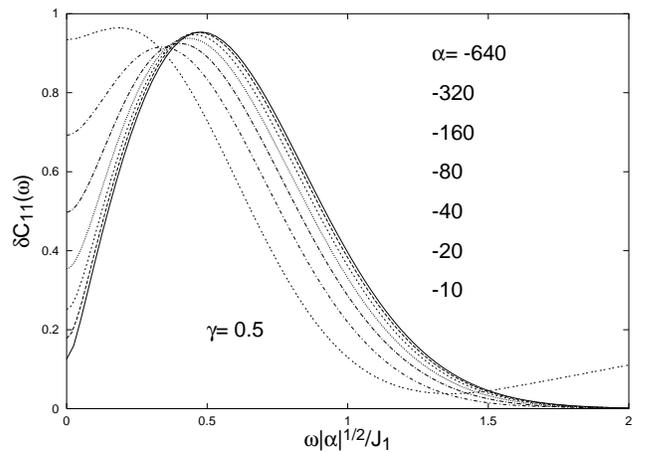}}
 \vspace{0.3cm}
 \caption{Low-$T$ plots of $\delta \tilde{\cal C}_{11}(\omega)$ vs.
  the scaled frequency $\omega |\alpha|^{1/2}/|J_{1}|$ displaying the
  AFM $\Omega_2(1/2)$ mode.}
 \label{akfig11}
\end{figure}

For the AFM case with $\gamma<1/2$, there are four modes at finite frequencies.
 For $1\ne\gamma>1/2$, $\Omega_1$ and $\Omega_4$ are
 degenerate, and the mode with frequency $\Omega_2(\gamma)=0$ 
is a central peak.  The two triangles in Fig.\ \ref{akfig10} 
indicate special points.
For $\gamma=1$,  the four  modes
 are all degenerate, $\delta\tilde{\cal
 C}_{11}(\omega)=\delta\tilde{\cal C}_{22}(\omega)$, and
 $\delta\tilde{\cal C}_{11}(\omega)$ approaches the AFM scaling
 form shown in Fig.\ \ref{akfig5}. 
 For the low-$T$ AFM case $\gamma=1/2$,   
$\delta\tilde{\cal C}_{11}(\omega)$ exhibits the mode frequency 
$\Omega_2(1/2)\rightarrow|J_2|/|\alpha|^{1/2}$  as
$T\rightarrow0$, and the  shape of this mode approaches a scaling function of 
$|\alpha|^{1/2}\omega/|J_1|$, as shown in 
 Fig.\ \ref{akfig11}. These  $\delta
 \tilde{\cal C}_{11}(\omega)$ curves scale {\it without} a
 corresponding rescaling of $\delta\tilde{\cal C}_{11}(\omega)$ itself.

\begin{figure}
 \epsfxsize=8.5cm
 \centerline{\epsffile{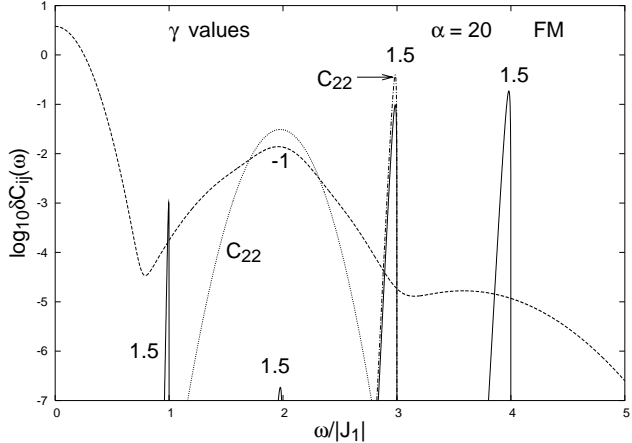}}
 \vspace{0.3cm}
 \caption{Plots at $\alpha=20$ of the FM ${\rm log}_{10}[\delta
  \tilde{\cal C}_{11}( \omega)]$ for $\gamma=1.5$ (solid) and -1 (dashed) and ${\rm log}_{10}[\delta
 \tilde{\cal C}_{22}
  (\omega)]$ for $\gamma=1.5$ (dot-dashed) and -1 (dotted)  
 versus $\omega/|J_1|$.}
 \label{akfig12}
\end{figure}

 We now consider the FM case at low $T$. To illustrate the types of 
 behavior found,
 in Fig.\ \ref{akfig12}, we  plot  
$\log_{10}[\delta\tilde{\cal C}_{11}(\omega)]$ and 
$\log_{10}[\delta\tilde{\cal C}_{22}(\omega)]$
 versus $\omega/|J_1|$ at $\alpha=20$ for  $\gamma=1.5, -1$. 
  For $\gamma=1.5$,
  $\delta\tilde{\cal C}_{11}(\omega)$ at this $T$ exhibits four sharp modes at 
$\omega/|J_1|\approx 1, 2, 3, 4$,
 respectively.  For $\gamma=-1$, there are three broad 
 $\delta\tilde{\cal C}_{11}(\omega)$ 
modes, with rather
 well-defined peaks at $\omega/|J_1|\approx 0, 2, 3.6$ at this $T$. As for the
 AFM case, in each case $\delta\tilde{\cal C}_{22}(\omega)$
is a single peak at one of the non-vanishing $\delta\tilde{\cal
 C}_{11}(\omega)$ peak positions.

\begin{figure}
 \epsfxsize=8.5cm
 \centerline{\epsffile{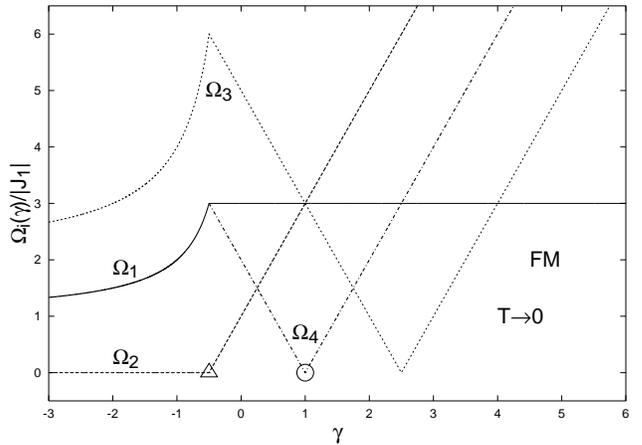}}
 \vspace{0.3cm}
 \caption{Plots of the low-$T$  FM mode frequencies $\Omega_i/J_{1}$ versus
 $\gamma$.  The triangle indicates low-$T$ mode scaling, and the circle 
indicates the disappearance of the mode.  See text.}
 \label{akfig13}
\end{figure}

 From an extensive analysis of many lower-$T$ results, we found that 
there are generally
 four low-$T$ $\delta\tilde{\cal C}_{11}(\omega)$ FM mode frequencies,
 $\Omega_i(\gamma)$, and that $\delta\tilde{\cal C}_{22}(\omega)$
 has one non-vanishing mode frequency, 
 $\Omega_1(\gamma)$. These FM $\Omega_i(\gamma)$, pictured in 
Fig.\ \ref{akfig13}, satisfy
\begin{eqnarray}
        \Omega_{1}(\gamma)/J_{1} &=& \left| 1-1/\gamma \right|, ~~
                       {\rm for}~~ \gamma \leq -1/2 , \nonumber \\
                   &=& 3,~~{\rm for}~~ \gamma \geq -1/2 ;
                \label{Om1FM} \\
        \Omega_{2}(\gamma)/J_{1} &=& 0,~~{\rm for}~~ \gamma \leq -1/2 ,
                       \nonumber \\
                   &=& 1+2\gamma ,~~{\rm for}~~ \gamma \geq -1/2 ;
                \label{Om2FM} \\
        \Omega_{3}(\gamma)/J_{1} &=& 2 \left| 1-1/\gamma \right|, ~~
                       {\rm for}~~ \gamma < -1/2 , \nonumber \\
                   &=& |5-2\gamma|,~~ {\rm for}~~ \gamma \geq -1/2 ;
                \label{Om3FM} \\
        \Omega_{4}(\gamma)/J_{1} &=&\left| 1-1/\gamma \right|, ~~
                       {\rm for}~~ \gamma \leq -1/2 , \nonumber \\
                   &=& 2|1-\gamma| ,~~{\rm for}~~ \gamma \geq -1/2 .
                \label{Om4FM}
\end{eqnarray}

 As for the AFM case with $1\ne\gamma>1/2$, for $\gamma<-1/2$,
 the low-$T$ FM mode with frequency $\Omega_2(\gamma)=0$ is a central  peak, 
and $\Omega_1$ and $\Omega_4$ are finite and degenerate. 
For $\gamma>-1/2$, there are four FM magnon modes, none of
 which is a central peak.  The amplitude of the weakest mode with frequency 
$\Omega_3(\gamma)$ is
  $\propto T^2$ as $T\rightarrow0$ for $\gamma\ge-1/2$.  At $\gamma=5/2$,
 $\Omega_3(5/2)\propto T$, so the mode with frequency $\Omega_3(\gamma)$ is 
never a central peak.  

 For the low-$T$ FM equilateral triangle, indicated by the circle in
 Fig.\ \ref{akfig13}, $\delta\tilde{\cal C}_{22}(\omega)=
\delta\tilde{\cal C}_{11}(\omega)$,
  since the  amplitude  of the mode with
 frequency $\Omega_4(\gamma)$ 
 vanishes as $\gamma\rightarrow1$,  
 and the remaining mode frequencies are
 degenerate.  Hence, 
 $\delta\tilde{\cal C}_{11}(\omega)$ 
 approaches the FM scaling form shown in Fig.\ \ref{akfig6}.

\begin{figure}
 \epsfxsize=8.5cm
 \centerline{\epsffile{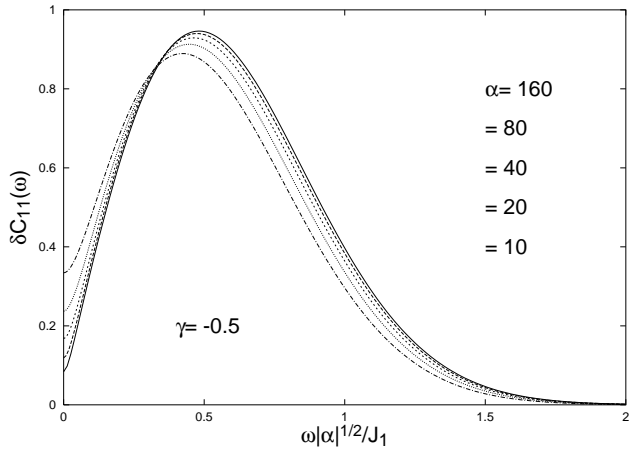}}
 \vspace{0.3cm}
 \caption{Plots at various low $T$ values of  $\delta
  \tilde{\cal C}_{11}(\omega)$ vs. the scaled frequency
  $\omega |\alpha|^{1/2}/|J_{1}|$ for the FM $\Omega_2(-0.5)$ mode.}
 \label{akfig14}
\end{figure}

 We now examine the other special FM case,  $\gamma=-1/2$. The mode $\Omega_2(-1/2)$, indicated
 by the triangle in Fig.\ \ref{akfig13}, is
  a  low-$T$ FM mode with scaling
 properties remarkably similar to those of the
AFM mode with frequency $\Omega_2(1/2)$, pictured in Fig.\ \ref{akfig10}.
 As $T\rightarrow0$,  the $\delta\tilde{\cal C}_{11}(\omega)$ FM mode
 frequency $\Omega_2(-1/2)\rightarrow|J_2|/|\alpha|^{1/2}$,
 and the mode shape approaches a scaling function of 
$|\alpha|^{1/2}\omega/|J_1|$,
 as pictured in  
 Fig.\
 \ref{akfig14}.
 By
 comparing Fig.\ \ref{akfig14} with Fig.\ \ref{akfig11},  the 
lowest $T$ curves in the two
 figures are almost identical, provided that one rescales $T$ by
 a factor of 8.  Since in  these special cases, the scaling  does not
 involve scaling of the mode amplitude,  it does not represent  a
 contribution to
 $\delta{\cal C}_{11}(t)$ involving a  scaling  of the time, as for the AFM equilateral triangle pictured in
 Fig.\ \ref{akfig5}, or more generally 
 for the AFM low-$T$ equivalent neighbor and four-spin ring models.
\cite{4sp,EqNeigh} It is therefore a new form of mode scaling, not previously 
 found in  any SMM system.

%
%

\section{Summary and Discussion}   \label{Sec:Conclusions}

 We presented the exact solution for the thermal equilibrium
 dynamics of three classical Heisenberg spins on a triangle with two exchange 
 couplings $J_1$ and $J_2$.   Although one might expect that for 
$\gamma=J_2/J_1\approx1$
  the behavior would not be too different from that of the equilateral
 triangle, $\gamma=1$, we found that this is not the case, regardless of $T$.  
 Instead of the two ${\cal C}_{ij}(t)$ related by a sum rule for $\gamma=1$, 
for $\gamma\ne1$ there are four ${\cal C}_{ij}(t)$ related by a sum
 rule. For $\gamma\ne0$,
 the piecewise continuous Fourier transforms $\delta\tilde{\cal C}_{ij}(\omega)$ demonstrate the
 profound effects of the absence of translational symmetry within the triangle 
 with $\gamma\ne1$.  As $T\rightarrow\infty$, they allow us to determine the
long-time behavior of the  autocorrelation functions ${\cal C}_{11}(t)$ 
and ${\cal C}_{22}(t)$, which
 approach 
 their distinct long-time asymptotic limits 
 differently,  as $t^{-2}$ and $t^{-3}$, respectively. For the three-spin
 chain, $\gamma=0$, the autocorrelation function on the chain end,
 $\delta\tilde{\cal C}_{11} (\omega)$, is discontinuous
 at $\omega/|J_1|=1$ at all $T$, leading to the
 characteristic $t^{-1}$ behavior of ${\cal C}_{11}(t)$ as 
$T\rightarrow\infty$.

 At low $T$, setting $\gamma\ne1$ leads to qualitatively
 different behavior  from that obtained for $\gamma=1$.  Regardless of the sign
 of $J_1$, for $\gamma\ne1$, there are four low-$T$  modes in
 $\delta\tilde{\cal C}_{11}(\omega)$.  
 For $J_1<0$ and $\gamma\ne1$, two of these low-$T$ modes, 
$\Omega_1$ and $\Omega_4$,
 are degenerate for $\gamma\ge1/2$.  
 For $J_1>0$, $\Omega_1$ and $\Omega_4$ are degenerate for $\gamma\le-1/2$.  
 In both of these FM and AFM regimes, one of the
 non-degenerate low-$T$ modes is a central peak with a width determined by the
 $J_i$,
  which grows in intensity as $T\rightarrow0$.  This central peak, which can
 have the largest mode intensity,
  arises from the
 pinning of a magnon mode in the isosceles triangle with $J_1\ne J_2$, and does
 not appear in any other model which has been solved exactly. 
 There is also a peak at $\omega=0$
 arising from the long-time asymptotic limit of  ${\cal
 C}_{11}(t)$, which would be broadened by relaxation processes not included in
 our model.     

For the three-spin chain, $\gamma=0$,
 the effects of the absence of translational symmetry in the chain 
are even more 
dramatic, leading to
 qualitatively different dynamical behaviors of the spins on the chain ends
 from that of the spin at the center.
 For both signs
 of $J_1$,
 $\delta\tilde{\cal C}_{22}(\omega)$ has only one mode energy $|J_1|$. But
 $\delta\tilde{\cal C}_{11}(\omega)$ is distinctly different.  For 
the AFM
 three-spin chain, $\delta\tilde{\cal C}_{11}(\omega)$ has three  
low-$T$ modes at $n|J_1|$ with $n= 1, 2, 3$,
 where the $n=1$ mode is doubly degenerate.  For the
 FM three-spin chain, $\delta\tilde{\cal C}_{11}(\omega)$ has four 
modes at $n|J_1|$ with $n=1, 2, 3, 5$.  Our
 results suggest that for  $N$-spin chains, the spins on or near the chain ends
 would have a 
richer dynamics than would those of more central spins, or of spins on closed
 rings of $N$ spins.
 It would be interesting to see if
 such differences are indeed maintained for larger spin chains and for
 larger spin clusters with two or more Heisenberg spin exchange interactions.

For the four-spin ring with equal near-neighbor exchange couplings
$J_1$, the single autocorrelation function $\delta\tilde{\cal C}_{11}(\omega)$
  has a single  mode that approaches the fixed frequency $2|J_1|$ as
$T\rightarrow0$, in addition to the AFM (FM) mode of the four-spin equivalent
 neighbor
model with a frequency that approaches 0 ($4|J_1|$). \cite{4sp,EqNeigh} None of these is a central
peak at $T\ne0$.
   The isosceles triangle has a much richer spectrum of four modes, 
with tunable frequencies that depend upon $\gamma$, and a central peak for a
semi-infinite range of $\gamma$ values.  In addition, at the onset of the central
peak, that mode form exhibits low-$T$ frequency scaling, instead of the 
time scaling
present in the AFM  equivalent neighbor and four-spin ring models, and it is a
scaling function of a differently scaled frequency than in the FM equivalent
neighbor model and four-spin ring.\cite{4sp,EqNeigh}
 The simplest integrable four-spin system
with dynamics similar to that of the isosceles triangle is the squashed
tetrahedron, which also involves two different near-neighbor exchange
couplings. 
\cite{SquashTet}

Inelastic neutron scattering experiments on large single crystals would be a
particularly useful technique to observe the effects predicted here.  By
appropriately varying the scattering wave vector, all four of the
$\delta\tilde{\cal C}_{ij}(\omega)$ can be measured as functions of $\omega$
and $T$.  
 Although experimental observation of the predicted low-$T$ modes might 
 be difficult, since quantum effects are expected to
 dominate at very low $T$, the presence of such modes for three classical
Heisenberg
spins on an isosceles triangle 
 underscores the qualitative changes that occur with different exchange
 couplings.  Our numerical results indicate that the development of these
 additional modes appears at $T$ values high enough for the
 classical treatment to be valid. 

%
%

\section*{Acknowledgments}
\noindent

 The authors thank P.\ Khalifah, K.\ Morawetz, S.\ E.\ Nagler, K.\ Scharnberg,
 and D.\ Sen for helpful discussions.

%
%

\section*{appendix}

 We first discuss the $T \rightarrow \infty$ limit of the integrals
 $I_{i}(t)$, Eqs.\ (\ref{i1})--(\ref{i3}). From Eq.\ (\ref{C22-I}),
 we obtain $\delta {\cal C}_{22}(t) = I_{1}(t)/4$. As $T \rightarrow
 \infty$,
 $\delta {\cal C}_{22}(t)$ coincides with that of the 
$N=3$ equivalent neighbor model,
 so $\lim_{T \rightarrow \infty} I_{1}(t)$ is obtained by
 setting $N=3$ in Eqs.\ (4)--(9) of Ref.\ \cite{EqNeigh}.

 Setting $t^{*}=J_{1}t$ as before, for $I_{2}(t)$ we find, 
\begin{equation}        \label{I2(t)}
        \lim_{T \rightarrow \infty} I_{2}(t) =
          \int_{0}^{2} dx f_{2}(x) \cos \left[ (1-\gamma) xt^{*}
          \right],
\end{equation}
\begin{equation}
        f_{2}(x) = \frac{4-x^{2}}{256 x^{2}} \left[ 4x (x^{2}+1)
          - (x^{2}-1)^{2} \ln \left( \frac{x+1}{x-1} \right)^{2}
          \right] .
\end{equation}
 Then Eq.\ 
 (\ref{C11longt}) is derived from $f_2(0)=f_2(2)=0$, and 
\begin{eqnarray}
        f'_{2}(0) &=& \frac{1}{6} , ~~ f'_{2}(2) = -\frac{5}{32}
          + \frac{9}{128} \ln 3 . \label{f2data}
\end{eqnarray}
 Since $f_{2}(x)$ and $f_2'(x)$ are continuous at  $x=1$,  
  Eq.\ (\ref{f2data}) leads to the  constants $a_1$ and $a_{2}$ in Eq.\
 (\ref{C11longt}). The
 FT of $I_3(t)$ at arbitrary $T$ is given in Eqs.\
 (\ref{FT_I3a})--(\ref{F113}).

 We illustrate the procedure for computing the FTs by obtaining
 $\delta \tilde{ {\cal C}}_{22}(\omega)$.  The FT  of $\cos (st^{*})$ 
 is $\delta
 (\tilde{\omega}+s) + \delta (\tilde{\omega}-s)$, where
 $\tilde{\omega}=\omega /J_{1}$. For $\tilde{\omega}>0$, the
 $\delta (\tilde{\omega}+s)$ term is irrelevant. Then $\tilde{\omega}$ 
satisfies
 $|x-1| \leq \tilde{\omega} \leq x+1$. For fixed $\tilde{\omega}$,
 the integration interval for $x$ is then determined as a function
 of $\tilde{\omega}$. Hence,
\begin{equation}
        \delta \tilde{{\cal C}}_{22}(\omega) = \frac{1}{4Z}
          \Theta (3-\tilde{\omega})
          \int_{|\tilde{\omega}-1|}^{{\rm min}[2,(\tilde{\omega}+1)]}
          dx {\cal F}_{22}(x),
          \label{C22ftcalc}
\end{equation}

\begin{equation}
        {\cal F}_{22}(x) = e^{\alpha [\tilde{\omega}^{2} + (\gamma -1)
          x^{2}]} \tilde{\omega} \left[ 1 - \left(
          \frac{\tilde{\omega}^{2}+1-x^{2}}{2\tilde{\omega}} \right)^{2}
          \right] ,
\end{equation}
 and $\Theta(z)$ is Heaviside's step function, $\Theta(z)=0$ if $z<0$,
 $\Theta(z)=1$ if $z>0$.

 The exact calculation of $\delta \tilde{{\cal C}}_{11}$ is substantially
 more involved. We define the functions $\tilde{I}_{a}(\omega)$ to be
 the FT's of the $I_{a}(t)$, Eqs.\ (\ref{i1})--(\ref{i3}),
\begin{equation}        \label{C11ft}
        \delta \tilde{{\cal C}}_{11}(\omega) =
          \tilde{I}_{1}(\omega) + \tilde{I}_{2}(\omega) +
          \tilde{I}_{3}(\omega) ,
\end{equation}
 where $\tilde{I}_{1}(\omega)= \delta \tilde{{\cal C}}_{22}(\omega)/4$.
 We then find that
\begin{eqnarray}
        \tilde{I}_{2}(\omega) &=& \frac{1}{2Z|1-\gamma|}
          \Theta (2 - \bar{\omega}) \int_{|1-\bar{\omega}|}^{1+\bar{\omega}}
          ds {\cal F}_{11,2}(s) , \label{I2FTint} \\
        {\cal F}_{11,2}(s) &=& e^{\alpha [s^{2} + (\gamma -1)
          \tilde{\omega}^{2}]} \times \nonumber \\
        & & \hspace{-10pt} \times
           s \left( 1-\frac{\bar{\omega}^{2}}{4} \right)
          \left[ 1 - \left( \frac{s^{2}+\bar{\omega}^{2}-1}{2s\bar{\omega}}
          \right)^{2} \right] ,
\end{eqnarray}
 and $\bar{\omega}=\omega/[J_{1}(1-\gamma)]$.  We first write
 $\tilde{I}_{3}(\omega)$   as
\begin{equation}
        \tilde{I}_{3}(\omega) = \frac{1}{16Z}
          \left[ \tilde{I}_{3a}(\tilde{\omega})
          + \tilde{I}_{3b}(\tilde{\omega})
          + \tilde{I}_{3c}(\tilde{\omega}) \right] .
\end{equation}
 For $0<\gamma<2$, setting $\Delta = |1-\gamma|$, we then have
\begin{eqnarray}
        \tilde{I}_{3a}(\tilde{\omega}) &=& \Theta (3+2\Delta-\tilde{\omega})
          \Theta (\tilde{\omega}-1) \times \nonumber \\
        & & \hspace{1pt} \times \int_{(\tilde{\omega}-1)/(1+\Delta)}^{
          {\rm min}[2, (\tilde{\omega}+1)/(1+\Delta)]} dx
          {\cal F}_{11,3-}(x) + \nonumber \\
        & & \hspace{-24pt}
          + \Theta (1-\tilde{\omega}) \Theta (\tilde{\omega}-\Delta)
          \times \nonumber \\
        & & \hspace{1pt} \times \int_{(1-\tilde{\omega})/(1-\Delta)}^{
          (1+\tilde{\omega})/(1+\Delta)} dx {\cal F}_{11,3-}(x),
        \label{FT_I3a} \\
        \tilde{I}_{3b}(\tilde{\omega}) &=& \Theta (3-2\Delta-\tilde{\omega})
          \Theta (\tilde{\omega}-1) \times \nonumber \\
        & & \hspace{1pt} \times \int_{(\tilde{\omega}-1)/(1-\Delta)}^{2}
          dx {\cal F}_{11,3+}(x) + \nonumber \\
        & & \hspace{-24pt}+ \Theta (1-\tilde{\omega})
          \int_{(1-\tilde{\omega})/(1+\Delta)}^{{\rm min}[2,
          (1+\tilde{\omega})/(1-\Delta)]}
          dx {\cal F}_{11,3+}(x) ,
        \label{FT_I3b} \\
        \tilde{I}_{3c}(\tilde{\omega}) &=& \Theta (\Delta -\tilde{\omega})
          \times \nonumber \\
        & & \hspace{1pt} \times
          \int_{(1+\tilde{\omega})/(1+\Delta)}^{{\rm min}[2,
          (1-\tilde{\omega})/(1-\Delta)]} dx
          [-{\cal F}_{11,3-}(x)],
        \label{FT_I3c} \\
        {\cal F}_{11,3\pm}(x) &=& e^{\alpha [(\tilde{\omega}\pm x\Delta)^{2}
          +(\gamma-1)x^{2}]} \times \nonumber \\
        & & \times
          \frac{(1-x^{2}/4)[(\tilde{\omega}\pm x(1+\Delta))^{2}-1]^{2}}{
          (\tilde{\omega}\pm x\Delta)x^{2}}.
        \label{F113}
\end{eqnarray}
 For other values of $\gamma$, we obtain similar expressions.

As $T\rightarrow\infty$, one has
\begin{equation}
        \lim_{T \rightarrow \infty} \tilde{I}_{2}(\omega) =
          \frac{1}{|1-\gamma|} \Theta(2-\bar{\omega})
          f_{2}(\bar{\omega}).
\end{equation}

 The general expression for ${\rm lim}_{T\rightarrow\infty}\tilde{I}_{3}(\omega)$ 
 is too
 complicated to be given here. For the chain,
\begin{eqnarray}
        \lim_{T \rightarrow \infty} \tilde{I}_{3}^{\gamma=0}(\omega) &=&
          \Theta (5-\tilde{\omega}) \Theta(\tilde{\omega}-3)
          \frac{1}{64} \tilde{I}_{3a}^{\gamma=0}(\tilde{\omega})
          \nonumber \\
        & & + \Theta (3-\tilde{\omega}) \Theta(\tilde{\omega}-1)
          \frac{1}{64} \tilde{I}_{3b}^{\gamma=0}(\tilde{\omega})
          \nonumber \\
        & & + \Theta (1-\tilde{\omega}) \frac{1}{64}
          \tilde{I}_{3c}^{\gamma=0}(\tilde{\omega}),
\end{eqnarray}
\begin{eqnarray}
        \tilde{I}_{3a}^{\gamma=0}(\tilde{\omega}) &=&
          \frac{(5-\tilde{\omega})(-24
          -195 \tilde{\omega} +229 \tilde{\omega}^{2}
          + 47 \tilde{\omega}^{3} + 7 \tilde{\omega}^{4})}{48
          \tilde{\omega}} \nonumber \\
        & & + \frac{(\tilde{\omega}^{2}-1)(1+7\tilde{\omega}^{2})}{
          \tilde{\omega}^{2}} \ln \frac{\tilde{\omega}-1}{4} \nonumber \\
        & & - \frac{(\tilde{\omega}^{2}-4)(\tilde{\omega}^{2}-1)^{2}}{
          4\tilde{\omega}^{2}} \ln \frac{\tilde{\omega}+1}{
          2(\tilde{\omega}-2)}, \\
        \tilde{I}_{3b}^{\gamma=0}(\tilde{\omega})
          &=& \frac{3 \tilde{\omega}^{4} +
          67 \tilde{\omega}^{2} -24}{6 \tilde{\omega}} \nonumber \\
        & & - \frac{(\tilde{\omega}^{2}-1)(\tilde{\omega}^{4} +
          23 \tilde{\omega}^{2} +8)}{4 \tilde{\omega}^{2}}
          \ln \frac{\tilde{\omega}+1}{\tilde{\omega}-1} , \\
        \tilde{I}_{3c}^{\gamma=0}(\tilde{\omega}) &=&
          \frac{7 \tilde{\omega}^{4}
          - 6 \tilde{\omega}^{2} +951}{24} \nonumber \\
        & & + \frac{(1-\tilde{\omega}^{2})(1+7\tilde{\omega}^{2})}{
          \tilde{\omega}^{2}} \ln \frac{1-\tilde{\omega}^{2}}{16} \nonumber \\
        & & + \frac{(4-\tilde{\omega}^{2})(1-\tilde{\omega}^{2})^{2}}{
          4 \tilde{\omega}^{2}} \ln \frac{4(4-\tilde{\omega}^{2})}{
          1-\tilde{\omega}^{2}}.
\end{eqnarray}

 The leading asymptotic behavior, Eq.\ (\ref{C11longtchain}), arises from
\begin{eqnarray}
        \tilde{I}^{\gamma=0}_{3a}(5) &=& 0,\\
        \tilde{I}^{\gamma=0}_{3a}(3) &=& \tilde{I}^{\gamma=0}_{3b}(3)
        =\frac{137}{3}
          -\frac{592}{9} \ln 2 ,\\
        \tilde{I}^{\gamma=0}_{3b}(1) &=& \frac{23}{3},~~
          \tilde{I}^{\gamma=0}_{3c}(1)=\frac{119}{3}, \\
        \tilde{I}^{\gamma=0}_{3c}(0) &=& \frac{315}{8}-33 \ln 2.
\end{eqnarray}

 We now derive the AFM  low-$T$ mode frequency $\Omega_2(\gamma)$. As
  $\alpha \rightarrow -\infty$,  the integrand in Eq.\
 (\ref{C22ftcalc}) becomes sharply peaked about some
 $\omega$-dependent value  $x_{0}$. We  then perform an
 asymptotic evaluation of the form,
\begin{equation}        \label{asymp}
        \int_{a}^{b} dx e^{h(x)} \approx
          e^{h(x_{0})} \sqrt{2\pi \left| \frac{d^{2}h}{dx^{2}}
          \right|_{x_{0}}^{-1}},
\end{equation}
\begin{eqnarray}
        h(x) &=& \alpha \left[ \tilde{\omega}^{2} + (\gamma-1)x^{2} \right]
          + \ln \tilde{\omega} + \nonumber \\
        & & \hspace{15pt} + \ln \left[ 1 -
          \frac{(1+\tilde{\omega}^{2}-x^{2})^{2}}{4\tilde{\omega}^{2}}
          \right] .
\end{eqnarray}

 Inside the integration region, the integrand has exactly one maximum
 at $x_0$.  For $\gamma >1$,  $x_{0}^{(\gamma>1)} = |1-\omega| + \epsilon$,
 while for $\gamma <1$,
 $x_{0}^{(\gamma<1)} = 1+\omega - \epsilon$,
 where $\epsilon>0$ is  ${\cal O}(\alpha^{-1})$. 
 We then evaluate the integral in Eq.  (\ref{asymp}). It is 
 maximal for $\omega=\Omega_{1}$, where
\begin{equation}
        \Omega_{1}(\gamma)/|J_{1}| = \left| 1 - 1/\gamma \right|.
\end{equation}
 For $\gamma<1/2$, this result gives a spurious maximum, since it
 would require $x>2$. Since the result of the asymptotic evaluation
 is a monotonically increasing function of $\omega$ for $\omega<1$,
 we therefore conclude that $\Omega_{1}=|J_{1}|$ for $\gamma<1/2$.
 The proof of Eq. (\ref{Om1AFM}) is thus complete.

%
%


\begin{thebibliography}{99}
\bibitem{note-Riemann}{
   This is known as the first Riemann-Lebesgue lemma.}
\bibitem{delta3}{
   Our definition of $\delta_{3}$ agrees with that in
   \cite{EqNeigh}, but differs by a
   factor of 2 from that in \cite{lubanAJP,4sp}.}
\end{thebibliography}
\end{document}